\documentclass[final,5p,times,twocolumn]{elsarticle}
 \biboptions{comma,sort&compress}
\usepackage{graphicx}
\usepackage{here}
\usepackage[dvips]{epsfig}

\def\beq{\begin{equation}}
\def\eeq{\end{equation}}
\def\bea{\begin{eqnarray}}
\def\eea{\end{eqnarray}}
\def\nin{\noindent}

\begin{document}

\begin{frontmatter}

\title{Recent $\tau$ physics at BaBar}
\address[label1]{Universit\`a di Pisa \& INFN, Pisa, Italy.}
 \author[label1]{B. Oberhof} 
 \author{\\on behalf of the BaBar collaboration.}
\ead{benjamin.oberhof@infn.it}
\begin{abstract}
\nin
We report some new results on $\tau$ decays obtained by the BaBar collaboration using 468 fb$^{-1}$ of $e^+ e^-$ 
collisions recorded at the PEP-II asymmetric collider at Stanford Linear Accelerator Center. First We will show the results 
for the branching fractions for the decay of the $\tau$ to a charged hadron and two $K_S ^0$, 
$\tau^- \rightarrow h^- K_S ^0 K_S ^0 (\pi^0) \nu_{\tau}$ and the branching fractions for high-multiplicity decays with 3 or 5 
charged particles in the final state, either pions or kaons. 
We will then show the results for the search of $2^{nd}$ class hadronic current decays involving $\eta'$ mesons 
and the invariant mass spectra for $\tau^- \rightarrow h^- h^+ h^- \nu_{\tau}$ decays, where $h=\pi, K$. 

\end{abstract}
\end{frontmatter}
\section{Introduction}
The decays of the $\tau$ lepton can be used as a high-precision probe of the Standard Model (SM) and various models of new physics. 
Last generation B-factories, due to the high luminosity an high $\tau^+ \tau^-$ pair production cross section, offer and ideal 
environment for these studies \cite{ronban}.

As first result we will present the measurements of the branching fractions of 
$\tau^- \rightarrow \pi^- K^0 _S K^0 _S (\pi^0) \nu_{\tau}$ decays and the first search for 
$\tau^- \rightarrow K^- K^0 _S K^0 _S (\pi^0) \nu_{\tau}$ decays. 
The first two of these decays represents a major background for the search of CP violation in the decay rate asymmetry of 
$\tau^- \rightarrow \pi^- K^0 _S \nu_{\tau}$ \cite{kos}, and, due to the large uncertainty on the branching fraction \cite{kos2}, 
it's precise determination is important for future experiments aiming to measure CP violation in $\tau$ decays. 

Study of the three- and five-prong, where prong means charged track, either pion or kaon, decay modes of the $\tau$ lepton, 
allows one to test the Standard Model and search for evidence of new physics \cite{hmtau}. 
We present measurements of the (resonant) $\tau^- \rightarrow \pi^- \pi^+ \pi^- \eta \nu_{\tau}$, 
$\tau^- \rightarrow \pi^- \pi^+ \pi^- \omega \nu_{\tau}$, 
$\tau^- \rightarrow \pi^- f_1 \nu_{\tau}$ branching fractions. For this purpose 
we use the primary decay modes of the $\eta$, $\omega(782)$, and $f_1(1258)$: 
$\eta \rightarrow \gamma \gamma$, $\eta \rightarrow \pi^+ \pi^- \pi^0$, $\eta \rightarrow 3 \pi^0$; 
$\omega \rightarrow \pi^+ \pi^- \pi^0$ and $f_1 \rightarrow 2\pi^+ 2\pi^-$, $f_1 \rightarrow \pi^+ \pi^- \eta$. 
We measure the branching fractions of the non-resonant decays, where 
the non-resonant category includes possible contributions from broad resonances. 
We present a new limit on the branching fractions of the second-class current decay 
$\tau^- \rightarrow \pi^- \eta'(958) \nu_{\tau}$, and the first limits on the allowed
first-class current decays $\tau^- \rightarrow K^- \eta'(958) \nu_{\tau}$ 
and $\tau^- \rightarrow \pi^- \pi^0 \eta'(958) \nu_{\tau}$. 
We set also the first limits on the branching fractions of five-prong decay modes in 
which one or more of the charged hadrons is a charged kaon. 
In all high multiplicity measurements we exclude the contribution of $K_S ^0\rightarrow \pi^+ \pi^-$ decays. 

Finally we present the results for the exclusive invariant mass distributions for the decays 
$\tau^- \rightarrow \pi^- \pi^+ \pi^- \nu_{\tau}$, $\tau^- \rightarrow K^- \pi^+ \pi^- \nu_{\tau}$, 
$\tau^- \rightarrow K^- K^+ \pi^- \nu_{\tau}$ and $\tau^- \rightarrow K^- K^+ K^- \nu_{\tau}$. 
These decays contain a rich and interesting spectrum of low energy QCD resonances and provide a 
clean environment to probe low energy QCD and measure fundamental properties of the Standard Model \cite{tauinvmcab}. 
The measurements of the strange spectral function obtained from $\tau$ lepton decays to final 
states containing kaons, for example, can be used for a combined fit of the strange quark mass, $m_S$, 
and the Cabibbo Kobayashi Maskawa (CKM) matrix element $|V_us|$ \cite{tauinvmstrange}. 
Recent measurements of these branching fractions and spectral functions, interpreted in the 
framework of the OPE and Finite Energy Sum Rules, suggest a value of $|V_{us}|$ that is approximately 
three standard deviations lower than Standard Model expectations from CKM unitarity \cite{tauinvmstrange}. 
For this analysis detector effects, in particular the resolution, scale and efficiency 
have been removed using Bayesian Unfolding \cite{tauinvmunf}. 
The decay structure for the $\tau^- \rightarrow h^- h^+ h^- \nu_{\tau}$ is shown both in the unfolded two 
particle invariant mass distributions and three dimensional distribution which is presented as Dalitz plots 
in slices of the three body invariant mass along with the projections. 
These distributions are of particular interest to model builders 
to study the rich decay structure of the $\tau$ lepton.

All this analyses are based on data recorded with the BaBar detector at the PEP-II asymmetric-energy $e^+e^-$ storage 
rings operated at the SLAC National Accelerator Laboratory. With an integrated luminosity $\mathcal L = 424 +  
44$ fb$^-1$ recorded at center-of-mass (CM) energies of $10.58$ GeV and $10.54$ GeV, respectively, and an averaged 
$\tau^+ \tau^-$ 
production cross section of $\sigma_{\tau \tau} = (0.919 \pm 0.003)$ nb \cite{ronban}, the data sample amounts to 
430 million $\tau$ pairs. 
The BaBar detector is described in detail in Ref. \cite{babar}. Charged-particle momenta are measured with a five-layer 
double-sided silicon vertex tracker and a 40-layer drift chamber, both operating in the 1.5 T magnetic field of a 
superconducting solenoid. Information from a detector of internally reflected Cerenkov light is used in conjunction 
with specific energy loss measurements from the tracking detectors to identify charged pions and kaons \cite{babardet}. 
Photons are reconstructed from energy clusters deposited in a CsI(TI) electromagnetic calorimeter. 
Electrons are identified by combining tracking and calorimeter information. 
An instrumented magnetic flux return is used to identify muons.
The background contamination and selection efficiencies are determined using Monte Carlo simulation. The 
$\tau$-pair production is simulated with the KK2F event generator \cite{kk2f}. The $\tau$ decays, continuum $q \bar q$,  events, 
and final-state radiative effects are modeled with the Tauola \cite{tauola}, JETSET \cite{jetset}, and Photos \cite{photos} 
generators, respectively. 
The detector response is simulated with GEANT4 \cite{geant}. 
All Monte Carlo events are processed through a full simulation of the BaBar detector and 
are reconstructed in the same manner as data. 

\nin
\section{The branching fraction of $\tau^- \rightarrow h^- K_S ^0 K_S ^0 (\pi^0) \nu_{\tau}$ decays}

The $\tau^- \rightarrow \pi^- K_S ^0 K_S ^0 \nu_{\tau}$ decay is simulated with Tauola 
using $\tau^- \rightarrow K^{*-} K^0 \nu_{\tau}$. 
The $\tau^- \rightarrow \pi^- K_S ^0 K_S ^0 \pi^0 \nu_{\tau}$ decay is simulated with EVTGEN using 
$\tau^- \rightarrow K^{*-} K^0 \pi^0 \nu_{\tau}$ and $\tau^- \rightarrow \pi^- K^{*0} K^0 \nu_{\tau}$. 
As we will see the $\tau^- \rightarrow K^{*-} K^0 \nu_{\tau}$ and 
$\tau^- \rightarrow K^{*-} K^0 \nu_{\tau}$ have a $K^*$(892) meson
that is observed in the $\pi^- K^0 _S$ channel, while the 
$\tau^- \rightarrow \pi^- K^{*0} K^0 \nu_{\tau}$  has a $K^{*0}$(892) meson that is observed
in $\pi^0 K^0 _S$ channel.
The decay products of the two $\tau$ leptons can be separated from each other by dividing the 
event into two hemispheres using the plane perpendicular to the event thrust axis [12]. 
The thrust axis is calculated using all charged particles and all neutral deposits event. 
We select events with one prompt track, with closest approach to the beam spot less than 1.5 cm 
in the plane transverse to the beam axis and 2.5 cm in the longitudinal direction,   
and two reconstructed $K^0 _S$ candidates in the signal hemisphere. 
In the other hemisphere we require one oppositely charged track. 

\begin{figure}[h] 
\begin{center}
\includegraphics[width=.4\textwidth]{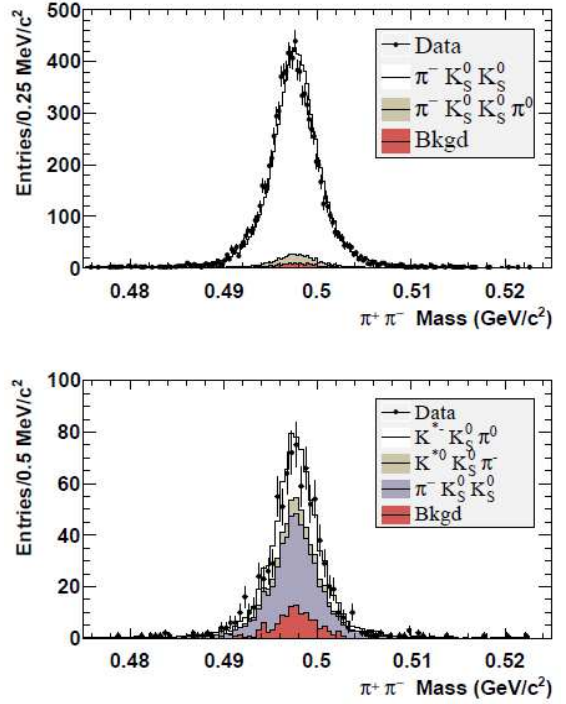}
\label{fig:kos_Spectra} 
\caption{\scriptsize The invariant mass of the two $K^0 _S \rightarrow \pi^+ \pi^-$ candidates
for the $\tau \rightarrow \pi K^0 _S K^0 _S \nu_{\tau}$ (top) and $\tau \rightarrow \pi K^0 _S K^0 _S \pi^0 \nu_{\tau}$
(bottom) samples after selection.
The points are data and the histograms are the prediction of the Monte Carlo simulation. For both plots,
the white histogram represents $\tau^- \rightarrow K^{*-} K^0 \nu_{\tau}$ decays, 
the blue and beige histogram shows the $\tau^- \rightarrow K^{*-} K^0 \pi^0 \nu_{\tau}$ and
$\tau^- \rightarrow K^{*0} K^0 \pi^- \nu_{\tau}$ decays, respectively while the red histogram is the $q \bar q$ background.}
\end{center}
\end{figure}

A $K^0 _S$ candidate is defined as a pair of oppositely charged tracks, 
with an invariant mass between 0.475 and 0.525 GeV/c$^2$ (fig. \ref{fig:kos_Spectra}). 

The charged pion and kaon samples are divided into samples with zero and one $\pi^0$ mesons. 
Events with two or more $\pi^0$ mesons are rejected. 
$\pi^0$ candidate are reconstructed from two clusters in the calorimeter, 
with a minimum energy of 30 MeV, and an invariant mass between 0.115 GeV/c$^2$ and 0.150 GeV/c$^2$. 
To reduce backgrounds from non-$\tau$-pair events, we require the momentum of the charged particle in the 
tag hemisphere to be less than 4 GeV/c in the CM frame and to be identified either as an electron or a muon. 

\begin{figure}[h] 
\begin{center}
\includegraphics[width=.4\textwidth]{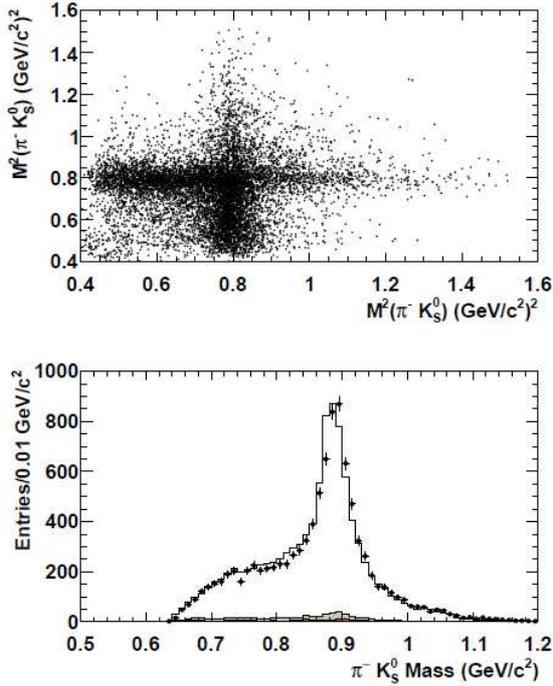}
\label{fig:kos3} 
\caption{\scriptsize The Dalitz plot and the invariant mass distributions for of the $K^0 _S \pi^-$ system. 
of selected events. There are two entries per event in the Dalitz plot and in the $K^0 _S \pi^-$ mass plot. 
The points are data and the histograms are the prediction of the Monte Carlo simulation. 
The white histogram represents $\tau^- \rightarrow K^{*-} K^0 \nu_{\tau}$ decays, 
the blue and beige histogram shows the $\tau^- \rightarrow K^{*-} K^0 \pi^0 \nu_{\tau}$ and
$\tau^- \rightarrow K^{*0} K^0 \pi^- \nu_{\tau}$ decays, respectively while the red histogram is the $q \bar q$ background.}
\end{center}
\end{figure}

\begin{figure}[h] 
\begin{center}
\includegraphics[width=.4\textwidth]{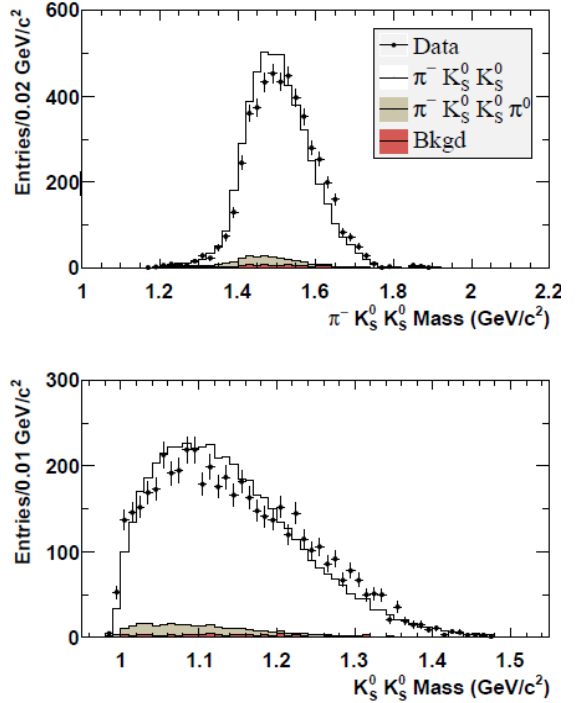}
\label{fig:kos4} 
\caption{\scriptsize Invariant mass distributions for events that pass the  selection criteria. 
The invariant mass requirement is not required for the plot of the $K^{0} _S K^0 _S \pi^-$ invariant mass. 
The points are data and the histograms are the prediction of the Monte Carlo simulation. 
The white histogram represents $\tau^- \rightarrow K^{*-} K^0 \nu_{\tau}$ decays, 
the blue and beige histogram shows the $\tau^- \rightarrow K^{*-} K^0 \pi^0 \nu_{\tau}$ and 
$\tau^- \rightarrow K^{*0} K^0 \pi^- \nu_{\tau}$ decays, respectively while the red histogram is the $q \bar q$ background. 
The mass plots use $\tau^- \rightarrow K^{0} _S K^0 _S \nu_{\tau}$ events that have been weighted based
on the Dalitz plot distributions.}
\end{center}
\end{figure}

The invariant mass of the charged hadron and the two $K^0 _S$ mesons is required to be less than 1.8 GeV/c$^2$. 
The $\pi^- K^0 _S K^0 _S$ invariant mass distributions are shown in fig. \ref{fig:kos4}. 
The invariant mass distribution predicted by the MC for the hadronic final state particles and for their combinations 
do not perfectly describe the data. In particular, the peak of the invariant mass distribution in
the MC is found to peak approximately 5\% lower than the peak observed in the data. 
To improve the modeling of the data we have weighted the $\tau^- \rightarrow \pi^- K^0 _S K^0 _S \nu_{\tau}$ in 
Tauola using the Dalitz plot distribution for the $K^0 _S \pi^-$ invariant mass of fig. \ref{fig:kos3}. The weighted 
events are used in all the mass plots and we observe an improvement in the modeling of the data. 
The branching fractions of the two charged pion modes are determined simultaneously to take into account the 
cross feed of each decay mode into the other sample.

Table \ref{tab:kos_res} summarizes the number of data and background events for each reconstruction mode as 
well as the selection efficiency matrix. 

We measure the $\tau^- \rightarrow \pi^- K^{0} _S K^0 _S \nu_{\tau}$ and 
$\tau^- \rightarrow \pi^- K^{0} _S K^0 _S \pi^0 \nu_{\tau}$ branching fractions to be:
\begin{equation}
B (\tau^- \rightarrow \pi^- K^{0} _S K^0 _S \nu_{\tau}) = (2.31 \pm 0.04 \pm 0.08) \times 10^{-4}
\end{equation}
and
\begin{equation}
B (\tau^- \rightarrow \pi^- K^{0} _S K^0 _S \pi^0 \nu_{\tau}) =(1.60 \pm 0.20 \pm 0.22) \times 10^{-5}
\end{equation}
respectively. 

\begin{table}[h]
\begin{center}
\begin{tabular}{lll}
& {\scriptsize $\tau^- \rightarrow \pi^- K^{0} _S K^0 _S \nu_{\tau}$} 
& {\scriptsize $\tau^- \rightarrow \pi^- K^{0} _S K^0 _S \pi^0 \nu_{\tau}$ }\\
\hline
{\scriptsize Branching Fraction $(10^{-5})$} & {\scriptsize $(23.1 \pm 0.4 \pm 0.8)$} 
& {\scriptsize $(1.60 \pm 0.20 \pm 0.22)$ } \\
\hline
{\scriptsize Data Events} & {\scriptsize 4985} & {\scriptsize 409}\\
\hline
{\scriptsize Estimated Background} & {\scriptsize 98 $\pm$ 17} & {\scriptsize 35 $\pm$ 7}\\
\hline
{\scriptsize Efficiency:} & & \\
{\scriptsize $\tau^- \rightarrow \pi^- K^{0} _S K^0 _S \nu_{\tau}$} & 
{\scriptsize $(4.93 \pm 0.02)$ \% } & {\scriptsize $(0.21 \pm 0.01)$ \% } \\
{\scriptsize $\tau^- \rightarrow \pi^- K^{0} _S K^0 _S \pi^0 \nu_{\tau}$} & 
{\scriptsize $(3.04 \pm 0.10)$ \% } & {\scriptsize $(2.65 \pm 0.09)$ \% } \\
\hline
{\scriptsize Selection Efficiency} & {\scriptsize 0.008} & {\scriptsize 0.12}\\
{\scriptsize Background} & {\scriptsize 0.004} & {\scriptsize 0.04}\\
{\scriptsize Common Contributions} & {\scriptsize 0.034} & {\scriptsize 0.03}\\
{\scriptsize Total} & {\scriptsize 0.035} & {\scriptsize 0.13}\\
\hline
\end{tabular}
\caption{\scriptsize Results for the charged pion decays including relative systematic contribution to the total error.}
\label{tab:kos_res}
\end{center}
\end{table}
\nin

The systematic uncertainties are divided into the selection efficiency, background, and common 
systematic components. The selection efficiencies include the MC statistical error and an error 
that takes into account the uncertainty for finding a fake $\pi^0$ meson. 
The background is predicted by the MC simulation to be entirely from 
$e^+ e^- \rightarrow q \bar q$ events and is confirmed with data and MC simulation control samples. 
The control samples are created using the nominal selection criteria except that the invariant mass requirements 
are reversed to eliminate the $\tau$ pair events and enhance $q \bar q$ events. 
The ratio of selected events in the data to MC control samples is found to be consistent
with unity within 15\% for both samples. The 15\% value is added to the MC statistical uncertainty 
of the number of background events. 
A number of systematic uncertainties are common to both branching fractions measurements. They can be categorized 
into two components: tracking and particle identification reconstruction uncertainties, and topological 
selection uncertainties. The tracking and particle identification reconstruction
uncertainties include the uncertainty on the track reconstruction 
efficiency (0.5\%) and the uncertainties on the efficiencies for particle identification: 
lepton identification (combined electron and muon) (1.6\%), charged pion particle identification
(0.5\%), and $K^0 _S$ identification (1.8\%). 
The topological selection uncertainties include a 2\% uncertainty associated with the selection criteria. 
The uncertainty on the number of $\tau$ pairs given by the product of the luminosity and 
the $e^+ e^- \rightarrow \tau^+ \tau^-$  cross section is also included (1\%).

The same criteria are used to select $\tau^- \rightarrow \pi^- K^{0} _S K^0 _S \nu_{\tau}$ 
and $\tau^- \rightarrow \pi^- K^{0} _S K^0 _S \pi^0 \nu_{\tau}$ decays except that the charged 
track is required to be a kaon. The numbers of events are given in table \ref{tab:kkos_res} and are found 
to be consistent with the estimated background prediction. The background is almost entirely due 
to cross feed of decays and very little contribution from $q \bar q$ events. 
The branching fractions are determined for each channel independently and used to place upper 
limits on the branching fractions of: 
\begin{equation} 
B (\tau^- \rightarrow \pi^- K^{0} _S K^0 _S \nu_{\tau}) < 6.3 \times 10^{-7}
\end{equation}
and
\begin{equation}
B (\tau^- \rightarrow \pi^- K^{0} _S K^0 _S \pi^0 \nu_{\tau}) < 4.0 \times 10^{-7}
\end{equation}
at the 90\% confidence level.

\begin{table}[h]
\begin{center}
\begin{tabular}{lll}
& {\scriptsize $\tau^- \rightarrow K^- K^{0} _S K^0 _S \nu_{\tau}$} 
& {\scriptsize $\tau^- \rightarrow K^- K^{0} _S K^0 _S \pi^0 \nu_{\tau}$ }\\
\hline
{\scriptsize Branching Fraction $(10^{-7})$} & {\scriptsize $(1.90 \pm 3.00 \pm 0.03)$} 
& {\scriptsize $(1.50 \pm 1.80 \pm 0.10)$ } \\
\hline
{\scriptsize Limit at 90\% CL} & {\scriptsize $6.3 \times 10^{-7}$} & {\scriptsize $4.0 \times 10^{-7}$ } \\
\hline
{\scriptsize Data Events} & {\scriptsize 23} & {\scriptsize 1}\\
\hline
{\scriptsize Estimated Background} & {\scriptsize 20 $\pm$ 0.50} & {\scriptsize 0.15 $\pm$ 0.02}\\
\hline
{\scriptsize Efficiency:} & {\scriptsize $(3.85 \pm 0.04)$ \% } & {\scriptsize $(1.37 \pm 0.03)$ \% } \\
\hline
\end{tabular}
\caption{\scriptsize Results for the charged kaon decays.}
\label{tab:kkos_res}
\end{center}
\end{table}

\nin



\section{High multiplicity $\tau$ decays}

We select events where one hemisphere (tag) contains exactly one track while the other hemisphere (signal) contains 
exactly three or five tracks with total charge opposite to that of the tag hemisphere. The event is rejected 
if any pair of oppositely charged tracks is consistent with a photon conversion. All tracks are required to have a point 
of closest approach to the interaction region less than 1.5 cm in the plane
transverse to the beam axis and less than 2.5 cm in the direction along that axis in order to reject 
tracks coming from $K^0 _S$ decays. To reduce backgrounds from non-$\tau$-pair events, we require that the momentum of the 
charged particle in the tag hemisphere be less than 4GeV/c in the CM frame and 
that the charged particle be identified as an electron or a muon. 
The $q \bar q$ background is suppressed by requiring that there be at most one energetic ($E > 1$GeV) electromagnetic 
calorimeter cluster in the tag hemisphere that is not associated with a track. Additional background 
suppression is achieved by requiring the magnitude of the event thrust to lie between 0.92 and 0.99. 
Neutral pion candidates are reconstructed in the signal hemisphere. If a photon candidate 
meets the invariant mass requirement with multiple photon candidates, then the neutral pion candidate 
with invariant mass closest to the nominal $\pi^0$ mass is selected. 
The search for additional pion candidates is then repeated using the remaining photon candidates. The 
residual photon clusters in the signal hemisphere are used to search for $\eta \rightarrow \gamma \gamma$ candidates. 
The branching fractions are calculated using the expression $B = N_S/(2 N_{\tau \tau} \epsilon)$ where $N_S$ is the number of candidates
after background subtraction, $N_{\tau\tau}$ is the number of $\tau$ pairs produced, and $\epsilon$ is the selection efficiency. 
The uncertainty of N is estimated to be 1\%. The selection efficiencies are determined from the signal Monte Carlo samples. 
The uncertainty on the selection efficiencies includes 0.5\% per 
track on the track reconstruction efficiency, as well as particle 
identification (PID) selection uncertainties. From 
studies conducted on real and simulated events, the uncertainties 
on the charged particle identification selectors 
are estimated to be 1\% for electrons, 2.5\% for muons, 
0.5\% for pions, and 1.8\% for kaons. The combined electron 
and muon particle identification uncertainty is estimated 
to be 1.6\% based on the composition of the event 
samples. The uncertainty on the $\pi^0 \rightarrow \gamma \gamma$ and $\eta \rightarrow \gamma \gamma$ 
reconstruction efficiency is estimated to be 3\% per candidate.

\subsection{Resonant high multiplicity $\tau$ decays} 

\subsubsection{$\tau^- \rightarrow 3 (\pi)^- \eta \nu_{\tau}$} 

\begin{figure}[h] 
\begin{center}
\includegraphics[width=.4\textwidth]{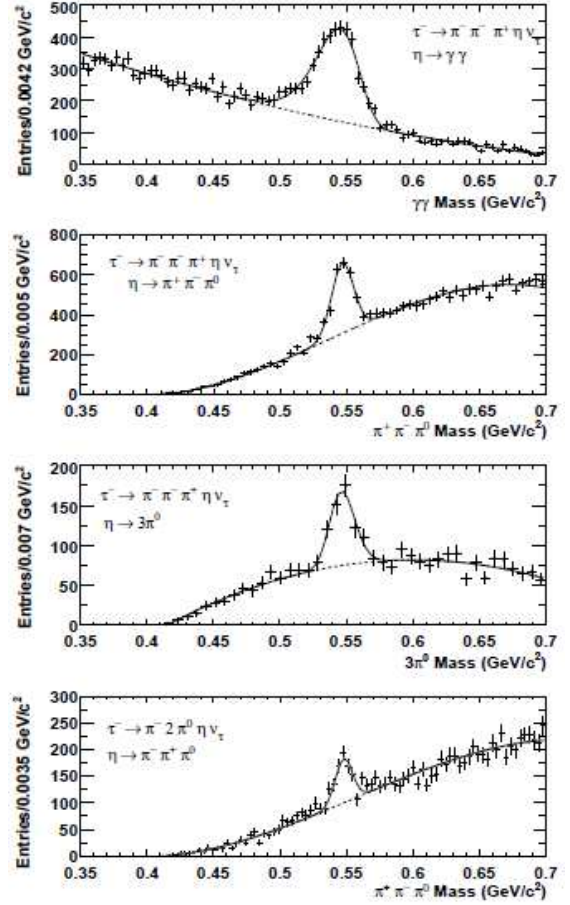}
\label{fig:eta} 
\caption{\scriptsize The $\gamma \gamma$, $\pi^+ \pi^- \pi^0$, and $3 \pi^0$ invariant mass distributions 
for $\tau^- \rightarrow \pi^- \pi^+ \pi^- \eta \nu_{\tau}$ candidates, and the $\pi^+ \pi^- \pi^0$ 
invariant mass distribution for $\tau^- \rightarrow \pi^- 2 \pi^0 \eta \nu_{\tau}$ decay candidates, 
after all selection criteria are applied. The solid lines represent the fit to the $\eta$ peak 
and background. The dashed lines show the extrapolation of the background function under the $\eta$ peak.}
\end{center}
\end{figure}

The $\tau^- \rightarrow \pi^- \pi^+ \pi^- \eta \nu_{\tau}$ mode is studied in the 
$\eta \rightarrow \gamma \gamma$, $\eta \rightarrow \pi^- \pi^+ \pi^0$, 
and $\eta \rightarrow 3 \pi^0$ final states, while the $\tau^- \rightarrow \pi^- 2 \pi^0 \eta \nu_{\tau}$ 
mode is studied in the $\eta \rightarrow \pi^- \pi^+ \pi^0$ final state. 
The event yields are determined by fitting the $\eta$ mass peak in the invariant mass distributions 
(fig. \ref{fig:eta}). The Monte Carlo simulation indicates that some of 
the entries in the peak are from $e^+ e^- \rightarrow q \bar q$ events.
Control samples, obtained by reversing the requirement 
on the invariant mass, are used to validate the background 
estimate. The expected background is corrected by the 
ratio of data to Monte Carlo events, and the statistical 
uncertainty of the ratio is included in the background 
systematic uncertainty. This method of validating the 
$q \bar q$ background estimate is used for all decays and is not 
mentioned in the later sections. 
The reconstruction efficiencies are determined from fits
to the signal Monte Carlo samples. 
The three determinations of the $\tau^- \rightarrow \pi^- \pi^+ \pi^- \eta \nu_{\tau}$ 
branching fraction are found to be in good agreement and we therefore calculate a weighted average. 
The weighted average is found to be 
\begin{equation}
B(\tau^- \rightarrow \pi^- \pi^+ \pi^- \eta \nu_{\tau}) = (2.25 \pm 0.07 \pm 0.12) \times 10^{-4}.
\end{equation}
The $\tau^- \rightarrow \pi^- 2 \pi^0 \eta \nu_{\tau}$ branching fraction, in turn, is found to be
\begin{equation}
B(\tau^- \rightarrow \pi^- 2 \pi^0 \eta \nu_{\tau}) = (2.01 \pm 0.34 \pm 0.22) \times 10^{-4}.
\end{equation}
Naively, the ratio of the $\tau^- \rightarrow \pi^- \pi^+ \pi^- \eta \nu_{\tau}$ to 
$\tau^- \rightarrow \pi^- 2 \pi^0 \eta \nu_{\tau}$ branching fractions is expected to be two if
the decay is dominated by the $\tau^- \rightarrow \pi^- f_1 \nu_{\tau}$ decay mode. The data do 
not support this expectation. The measurements are in good agreement with the results from 
the CLEO Collaboration, $(2.3 \pm 0.5) \times 10^{-4}$ and $(1.5 \pm 0.5) \times 10^{-4}$, for 
$\tau^- \rightarrow \pi^- \pi^+ \pi^- \eta \nu_{\tau}$ and $\tau^- \rightarrow \pi^- 2 \pi^0 \eta \nu_{\tau}$ 
respectively \cite{cleoeta}. 

Table \ref{tab:eta_res} summarizes the results for the individual channels. 

\begin{table}[h]
\begin{center}
\begin{tabular}{ccccc}
& {\scriptsize BF  in $10^{-4}$} & {\scriptsize $N_{S}$} & {\scriptsize $N_{bkg}$} & 
{\scriptsize $\epsilon$ (\%)} \\
\hline
{\scriptsize $\tau^- \rightarrow \pi^- \pi^+ \pi^- \eta \nu_{\tau}$}\\
{\scriptsize $\eta \rightarrow \gamma \gamma$} & 
{\scriptsize $2.10 \pm 0.09 \pm 0.13$} & {\scriptsize $2887 \pm 103$} & {\scriptsize $131 \pm 29$} 
& {\scriptsize $3.83 \pm 0.11$} \\
\hline
{\scriptsize $\tau^- \rightarrow \pi^- \pi^+ \pi^- \eta \nu_{\tau}$}\\
{\scriptsize $\eta \rightarrow \pi^- \pi^+ \pi^0$} & 
{\scriptsize $2.37 \pm 0.12 \pm 0.18$} & {\scriptsize $1440 \pm 68$} & {\scriptsize $65 \pm 38$} 
& {\scriptsize $2.97 \pm 0.12$} \\
\hline
{\scriptsize $\tau^- \rightarrow \pi^- \pi^+ \pi^- \eta \nu_{\tau}$}\\
{\scriptsize $\eta \rightarrow 3 \pi^0$} & 
{\scriptsize $2.54 \pm 0.27 \pm 0.25$} & {\scriptsize $315 \pm 34$} & {\scriptsize $13 \pm 7$} 
& {\scriptsize $0.42 \pm 0.01$} \\
\hline
{\scriptsize $\tau^- \rightarrow \pi^- \pi^+ \pi^- \eta \nu_{\tau}$} & 
{\scriptsize $2.01 \pm 0.34 \pm 0.22$} & {\scriptsize $381 \pm 45$} & {\scriptsize $83 \pm 12$} 
& {\scriptsize $05.7 \pm 0.02$} \\
\hline
\end{tabular}
\caption{\scriptsize Results and branching fractions for $\tau^- \rightarrow (3\pi)^- \eta \nu_{\tau}$ decays.}
\label{tab:eta_res}
\end{center}
\end{table}
\nin

\subsubsection{$\tau^- \rightarrow \pi^- f_1 \nu_{\tau}$} 

The branching fraction of $\tau^- \rightarrow \pi^- f_1 \nu_{\tau}$ as well as the $f_1$ mass 
are measured using the $f_1 \rightarrow 2\pi^+ 2\pi^-$ and $f_1 \rightarrow \pi^+ \pi^- \eta$ decay modes, 
where the last one is reconstructed using $\eta \rightarrow \gamma \gamma$, 
$\eta \rightarrow \pi^+ \pi^- \pi^0$, and $\eta \rightarrow 3 \pi^0$
events. 
With respect to the selection criteria already described we modify the selection for the mass
measurement, dropping the requirement that the track in the tag hemisphere be a lepton and the restriction on 
the number of photon candidates in the tag hemisphere, in order to increase the size of the event sample. 
The numbers of $\tau^- \rightarrow \pi^- f_1 \nu_{\tau}$ candidates are determined
by fitting the $f_1$ peak in the $2\pi^+ 2\pi^-$ and $\pi^+ \pi^- \pi^0$ 
invariant mass distributions. The $f_1$ lineshape is expected to be a Breit-Wigner distribution, modified 
by the limited phase space. Previous studies show that the $f_1 \rightarrow a_0^- \pi^+$, $(a_0^- \rightarrow \pi^- \eta$ 
channel accounts for all $f_1 \rightarrow \pi^+ \pi^- \eta$ decays \cite{f1_decay}. The mass of the 
$\pi^+ a_0^-(980)$ system and the $\tau$ mass provide a lower and upper limit, respectively, on the $f_1$ lineshape. 
We use EVTGEN generator to determine the simulated $f_1$ lineshape and find it to be a close approximation to the
Breit-Wigner expectation. The $f_1$ peak is fit using this lineshape convolved with a Gaussian distribution to take 
into account the effects of the detector resolution. 

The product of the $\tau^- \rightarrow \pi^- f_1 \nu_{\tau}$ and $f_1 \rightarrow 2\pi^+ 2\pi^-$ 
branching fractions, and the product of the $\tau^- \rightarrow \pi^- f_1 \nu_{\tau}$ 
and $f_1 \rightarrow \pi^+ \pi^- \eta$ branching fractions, are measured to be 
\begin{eqnarray}
& B(\tau^- \rightarrow \pi^- f_1 \nu_{\tau}) B(f_1 \rightarrow 2\pi^+ 2\pi^-) = \\ 
& (5.20 \pm 0.31 \pm 0.37) \times 10^{-5}\\
& B(\tau^- \rightarrow \pi^- f_1 \nu_{\tau}) B(f_1 \rightarrow \pi^+ \pi^- \eta) = \\
& (1.26 \pm 0.06 \pm 0.06) \times 10^{-4}
\end{eqnarray}
respectively, where the second result is the weighted average of the three $\eta$ modes. 
The $B(\tau^- \rightarrow \pi^- f_1 \nu_{\tau})$ branching fraction is determined to be $(4.73\pm0.28\pm0.45) \times10^{-4}$
and $(3.60 \pm 0.18 \pm 0.23) \times 10^{-4}$, as obtained by dividing the product branching fractions by 
$B(f_1 \rightarrow 2\pi^+ 2\pi^-) = 0.110^{+0.007}_{-0.006}$ and 
$B(f_1 \rightarrow \pi^+ \pi^- \eta) = 0.349^{+0.013}_{-0.015}$ \cite{f1_ratio}, respectively.

Table \ref{tab:f1_res} summarizes the results for the individual channels. 

\begin{table}[h]
\begin{center}
\begin{tabular}{cccc}
& {\scriptsize BF  in $10^{-4}$} & {\scriptsize $N_{S}$} & 
{\scriptsize $\epsilon$ (\%)} \\
\hline
{\scriptsize $f_1 \rightarrow 2 \pi^+ 2 \pi^-$} & 
{\scriptsize $0.520 \pm 0.031 \pm 0.037$} & {\scriptsize $3722 \pm 222$} 
& {\scriptsize $8.3 \pm 0.1$} \\
\hline
{\scriptsize $f_1 \rightarrow \pi^+ \pi^- \eta$}\\
{\scriptsize $\eta \rightarrow \gamma \gamma$} & 
{\scriptsize $1.25 \pm 0.08 \pm 0.07$} & {\scriptsize $1605 \pm 94$} 
& {\scriptsize $2.97 \pm 0.12$} \\
\hline
{\scriptsize $f_1 \rightarrow \pi^+ \pi^- \eta$}\\
{\scriptsize $\eta \rightarrow \pi^- \pi^+ \pi^0$} & 
{\scriptsize $1.26 \pm 0.11 \pm 0.08$} & {\scriptsize $731 \pm 62$} 
& {\scriptsize $2.97 \pm 0.05$} \\
\hline
{\scriptsize $f_1 \rightarrow \pi^+ \pi^- \eta$}\\
{\scriptsize $\eta \rightarrow 3 \pi^0$} & 
{\scriptsize $1.33 \pm 0.34 \pm 0.39$} & {\scriptsize $197 \pm 59$} 
& {\scriptsize $0.53 \pm 0.06$} \\
\hline
\end{tabular}
\caption{\scriptsize Results and branching fractions for $\tau^- \rightarrow \pi^- f_1 \nu_{\tau}$ decays.}
\label{tab:f1_res}
\end{center}
\end{table}
\nin

\begin{figure}[h] 
\begin{center}
\includegraphics[width=.4\textwidth]{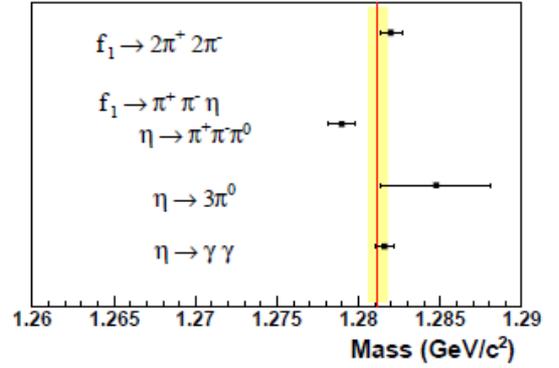}
\label{fig:f1_mass} 
\caption{\scriptsize Compilation of our measurements of the $f_1$ mass. 
The solid line is the weighted average and the shaded area is 
the one-standard-deviation region.}
\end{center}
\end{figure}

The $f_1$ mass is determined by fitting the peak with a non-relativistic Breit-Wigner function, which was used in
previous measurements of the $f_1$ mass \cite{f1_ratio}. As a cross check, we use the energy-momentum four-vectors from
the generator Monte Carlo simulation, and we find the fitted mass value to be consistent with the input mass value. 
The results for the different channels are shown in fig. \ref{fig:f1_mass}.

We determine the mass of the $f_1(1258)$ meson to be 
\begin{equation}
M_{f_1} = (1.28116 \pm 0.00039 \pm 0.00045) \mbox{ GeV/c}^2.
\end{equation}
The systematic uncertainty includes the reconstruction uncertainty and the calibration uncertainty. 
This result is in good agreement with the value $(1.2818 \pm 0.0006)$ GeV/c$^2$ in \cite{f1_ratio}.

\subsubsection{$\tau^- \rightarrow 3 (\pi)^- \omega \nu_{\tau}$} 

We measure the $\tau^- \rightarrow \pi^+ 2 \pi^- \omega \nu_{\tau}$ and 
$\tau^- \rightarrow \pi^- 2 \pi^0 \omega \nu_{\tau}$ branching fractions. 
The number of events is determined by fitting the $\eta$ peak in the $\pi^+\pi^-\pi^0$ invariant mass distributions
with a Breit-Wigner distribution, which is convolved with a Gaussian distribution to take
into account the detector resolution. A polynomial function is used to fit the background. 
The results are presented in \ref{tab:omega_res}. 
Approximately 10\% of the events in the $\tau^- \rightarrow \pi^+ 2 \pi^- \omega \nu_{\tau}$ channel are backgrounds 
from other $\tau$ decays (primarily $\tau^- \rightarrow \pi^- \pi^0 \omega \nu_{\tau}$ decays) 
and $e^+ e^- \rightarrow q \bar q$ events. 
The backgrounds are subtracted before calculating the branching fraction.
The $\tau^- \rightarrow \pi^- 2 \pi^0 \omega \nu_{\tau}$ sample has substantial contributions 
from $\tau^- \rightarrow \pi^- \omega \nu_{\tau}$ and 
$\tau^- \rightarrow \pi^- \pi^0 \omega \nu_{\tau}$ decays. The background is estimated with the Monte Carlo simulation
and verified using data and simulation control samples. The control samples follow the nominal selection criteria
but select one or two $\pi^0$ instead of three $\pi^0$ mesons.
The branching fractions are found to be
\begin{eqnarray}
B(\tau^- \rightarrow \pi^+ 2 \pi^- \omega \nu_{\tau}) & = & (8.4 \pm 0.4 \pm 0.6) \times 10^{-5}\\
B(\tau^- \rightarrow \pi^- 2 \pi^0 \omega \nu_{\tau}) & = & (7.3 \pm 1.2 \pm 1.2) \times 10^{-5}. 
\end{eqnarray}

\begin{table}[h]
\begin{center}
\begin{tabular}{lll}
& {\scriptsize $\tau^- \rightarrow \pi^+ 2 \pi^- \omega \nu_{\tau}$}
& {\scriptsize $\tau^- \rightarrow \pi^- 2 \pi^0 \omega \nu_{\tau}$}\\
\hline
{\scriptsize Branching Fraction} & {\scriptsize $(8.4 \pm 0.4 \pm 0.6) \times 10^{-5}$} 
& {\scriptsize $(7.3 \pm 1.2 \pm 1.2) \times 10^{-5}$ } \\
\hline
{\scriptsize Data Events} & {\scriptsize 2372 $\pm$ 94} & {\scriptsize 1135 $\pm$ 70}\\
\hline
{\scriptsize Estimated Background} & {\scriptsize 257 $\pm$ 71} & {\scriptsize 709 $\pm$ 59}\\
\hline
{\scriptsize Efficiency:} & {\scriptsize $(3.27 \pm 0.03)$ \% } & {\scriptsize $(0.75 \pm 0.01)$ \% } \\
\hline
\end{tabular}
\caption{\scriptsize Results and branching fractions for $\tau^- \rightarrow 3(\pi)^- \omega \nu_{\tau}$ decays.}
\label{tab:omega_res}
\end{center}
\end{table}

\subsection{Nonresonant high multiplicity $\tau$ decays}

The resonant modes, involving $\eta$, $\omega$ and $f_1$ mesons, do not account for all of the observed decays. 
We consider the excess in the observed decays to be from "non-resonant" modes. 
We make no attempt to identify the contribution of resonances with widths larger than 100 MeV/c$^2$ 
as the nature of these resonances is complex and their lineshapes will be modified
by the limited phase space. 
We measure the branching fractions of the non resonant $\tau^- \rightarrow \pi^+ 2 \pi^- 3\pi^0 \nu_{\tau}$, 
$\tau^- \rightarrow 2\pi^+ 3\pi^- \nu_{\tau}$ and $\tau^- \rightarrow 2\pi^+ 3\pi^- \pi^0 \nu_{\tau}$ decays. 
The numbers of candidates are given by the numbers of events found in the data after
subtracting the resonant contributions and the background from other $\tau$ decays and $q \bar q$ events. 
The resonant decays dominate the $\tau^- \rightarrow \pi^+ 2 \pi^- 3\pi^0 \nu_{\tau}$ mode.
The background is primarily from $\tau^- \rightarrow \pi^- \pi^0 \omega \nu_{\tau}$ and
$q \bar q$ events. The branching fraction of this non-resonant mode is determined to be 
\begin{equation}
B(\tau^- \rightarrow \pi^+ 2 \pi^- 3\pi^0 \nu_{\tau})=(1.0 \pm 0.8 \pm 3.0) \times 10^{-5}.
\end{equation}
The systematic uncertainty on the branching fraction is dominated by the uncertainty in the 
background, which includes the Monte Carlo statistical uncertainty and the $\tau$ branching fraction uncertainties. 
The branching fraction is consistent with zero and we set a limit of 
\begin{equation}
B(\tau^- \rightarrow \pi^+ 2 \pi^- 3\pi^0 \nu_{\tau}) < 5.8 \times 10^{-5}.
\end{equation}
at 90\% confidence level. We also determine the inclusive $\tau^- \rightarrow \pi^+ 2 \pi^- 3\pi^0 \nu_{\tau}$ 
branching fraction, given by the sum of the resonant and non-resonant terms. 
We find the result 
\begin{equation}
B(\tau^- \rightarrow \pi^+ 2 \pi^- 3\pi^0 \nu_{\tau})=(2.07 \pm 0.18 \pm 0.37) \times 10^{-4}
\end{equation}
where the systematic uncertainty accounts for correlations between the systematic uncertainties of 
the individual modes.

The $\tau^- \rightarrow 2\pi^+ 3\pi^- \nu_{\tau}$ decay has only a small contribution 
from resonant decays. The branching fraction is determined to be 
\begin{equation}
B(\tau^- \rightarrow 2\pi^+ 3\pi^- \nu_{\tau}) = (7.68 \pm 0.04 \pm 0.40) \times 10^{-4}.
\end{equation}
Although the modeling of the $2\pi^+ 3\pi^-$ invariant mass distribution is deficient, the
selection efficiency remains the same if the Monte Carlo is re-weighted to resemble the data distribution. 
The inclusive branching fraction is 
\begin{equation}
B(\tau^- \rightarrow 2\pi^+ 3\pi^- \nu_{\tau})=(8.33 \pm 0.04 \pm 0.43) \times 10^{-4}
\end{equation} and is obtained
by adding the non-resonant branching fraction with the resonant branching fraction 
for the $\tau^- \rightarrow \pi^- f_1 \nu_{\tau}$ via $f_1 \rightarrow 2\pi^+2\pi^-$ decay. 

$\tau^- \rightarrow 2\pi^+ 3\pi^- \pi^0 \nu_{\tau}$ decays are dominated by the resonant€
modes. We determine the branching fraction of the non-resonant $\tau^- \rightarrow 2\pi^+ 3\pi^- \pi^0 \nu_{\tau}$ 
decay to be 
\begin{equation}
B(\tau^- \rightarrow 2\pi^+ 3\pi^- \pi^0 \nu_{\tau}) = (3.6 \pm 0.3 \pm 0.9) \times 10^{-5}.
\end{equation}
The systematic uncertainty on this non-resonant branching fraction is dominated by the 
large uncertainty in the background. Although the invariant mass distributions of the resonant 
modes in the Monte Carlo simulation are corrected to provide better agreement with the data, 
the corrections make little difference to the final branching fraction result. 
The other $\tau$ decays and the $q \bar q$ events contribute to a lesser extent; their contribution 
to the uncertainty of the background is very small. 
The inclusive branching fraction
\begin{equation}
B(\tau^- \rightarrow 2\pi^+ 3\pi^- \pi^0 \nu_{\tau} = (1.65 \pm 0.05 \pm 0.09) \times 10^{-4} 
\end{equation}
and is obtained by adding the non-resonant branching fraction 
and the resonant branching fractions attributed to the 
$\tau \rightarrow 2\pi^- \pi^+ \eta \nu_{\tau}$ via $\eta \rightarrow \pi^+ \pi^- \pi^0$ 
and $\tau \rightarrow 2\pi^- \pi^+ \omega \nu_{\tau}$ via $\omega \rightarrow \pi^+ \pi^- \pi^0$ decays.

\subsection{5-prong $\tau$ decays with kaons}

We present here also the first search for high-multiplicity 
$\tau$  decays with one or two charged kaons. We find no evidence 
for signal decays and place upper limits on
the branching fractions of 5 different decay modes. 
The events are divided into topologies in which the
charged kaon has either the same or opposite charge as
the parent $\tau$ lepton. If there are two kaon candidates, 
they must have opposite charge. All other tracks are required 
to be identified as charged pions. 
Figure 9 shows the mass spectra 
for the various channels. The predictions of the Monte 
Carlo simulation are divided into decays with or without 
a K. The background estimates, which 
give the dominant systematic uncertainty, are verified by 
comparing the numbers of events in the data and Monte 
Carlo samples in the $M > 1.8$ GeV/c$^2$ region. 
The backgrounds
predicted by the Monte Carlo simulation are
approximately equal to the numbers of events in the
data sample. 
The upper limits on the branching fractions are given
in \ref{tab:res_k}. 

\begin{table}[h]
\begin{center}
\begin{tabular}{ccccc}
& {\scriptsize Limit $10^{-6}$} & {\scriptsize $N_{S}$} & {\scriptsize $N_{bkg}$} & 
{\scriptsize $\epsilon$ (\%)}  \\
\hline
{\scriptsize $\tau^- \rightarrow K^- 2\pi^- 2\pi^+ \nu_{\tau}$} & 
{\scriptsize $2.4$} & {\scriptsize $1328 \pm 36$} & {\scriptsize $1284 \pm 72$} 
& {\scriptsize $7.9 \pm 0.1$} \\
\hline
{\scriptsize $\tau^- \rightarrow K^+ 3\pi^- \pi^+ \nu_{\tau}$} & 
{\scriptsize $5$} & {\scriptsize $1999 \pm 45$} & {\scriptsize $1890 \pm 163$} 
& {\scriptsize $7.9 \pm 0.1$} \\
\hline
{\scriptsize $\tau^- \rightarrow K^- K^+ 2\pi^- \pi^+ \nu_{\tau}$} & 
{\scriptsize $0.45$} & {\scriptsize $32 \pm 6$} & {\scriptsize $15 \pm 4$} 
& {\scriptsize $6.7 \pm 0.1$} \\
\hline
{\scriptsize $\tau^- \rightarrow K^- 2\pi^- 2\pi^+ \pi^0 \nu_{\tau}$} & 
{\scriptsize $1.9$} & {\scriptsize $112 \pm 11$} & {\scriptsize $84 \pm 10$} 
& {\scriptsize $2.9 \pm 0.06$} \\
\hline
{\scriptsize $\tau^- \rightarrow K^+ 3\pi^- \pi^+ \pi^0 \nu_{\tau}$} & 
{\scriptsize $0.8$} & {\scriptsize $154 \pm 12$} & {\scriptsize $170 \pm 16$} 
& {\scriptsize $2.9 \pm 0.06$} \\
\hline
\end{tabular}
\caption{\scriptsize Upper limits at 90\% CL for charged kaon decay modes.}
\label{tab:res_k}
\end{center}
\end{table}

Currently there are no theoretical predictions for these modes. 
We estimate that $B(\tau^- \rightarrow K^- 2\pi^- 2\pi^+ \nu_{\tau} ) \sim 10^{-5}-10^{-6}$ 
if the decay is related to $B(\tau^- \rightarrow 3\pi^- 2\pi^+ \nu_{\tau})$ by the ratio
of the CKM matrix elements $(V_{us}/V_{ud})$.

\section{Search for $2^{nd}$ class current decays}

We show here the results search for the $\tau^- \rightarrow \pi^- \pi^0 \eta'(958) \nu_{\tau}$, 
$\tau^- \rightarrow K^- \eta'(958) \nu_{\tau}$, and $\tau^- \rightarrow \pi^- \eta'(958) \nu_{\tau}$ decays, where
$\eta' \rightarrow \pi^+ \pi^- \eta$. The first two decays are allowed first-class current 
decays whereas the last decay is a second-class current decay, wich rate would be zero in the limit
of perfect isospin symmetry. The event selection is similar to the previous sections. 
For the $\tau^- \rightarrow \pi^- \pi^0 \eta'(958) \nu_{\tau}$ via $\eta \rightarrow \gamma \gamma$ and 
the $\tau^- \rightarrow \pi^- \eta'(958) \nu_{\tau}$ via $\eta \rightarrow \gamma \gamma$ and 
$\eta \rightarrow \pi^+ \pi^- \pi^0$ modes, we 
measure the number of $\eta'$ candidates by fitting the peak 
with a Gaussian function and the combinatoric background
with a polynomial function. The number of $\eta'$ 
candidates in the other channels is determined by counting 
the number of events in a single bin centered on the 
$\eta'$ mass and subtracting the combinatorial events. The 
level of the combinatorial background is estimated by fitting 
the mass spectrum or from the average level of the 
sideband region around the $\eta'$ peak. 

We find no evidence for any of these decays and place the following upper limits on the 
branching fractions at the 90\% confidence level: 
\begin{eqnarray}
B(\tau^- \rightarrow \pi^- \pi^0 \eta'(958) \nu_{\tau}) & < & 1.2 \times 10^{-5}\\
B(\tau^- \rightarrow K^- \eta'(958) \nu_{\tau}) & < & 2.4 \times 10^{-6}\\
B(\tau^- \rightarrow \pi^- \eta'(958) \nu_{\tau}) & < &4.0 \times 10^{-6}.
\end{eqnarray}
The limits are determined from the weighted average of the branching fractions measured for each mode. 
The $\tau^- \rightarrow \pi^- \pi^0 \eta'(958) \nu_{\tau}$ and $\tau^- \rightarrow K^- \eta'(958) \nu_{\tau}$ 
channels are potential backgrounds to the $\tau^- \rightarrow \pi^- \eta'(958) \nu_{\tau}$ 
decay. We find that background from these two decays is less than two events based on the upper limits 
on the branching fractions and we consider these backgrounds to be negligible.  
It is predicted that the branching fraction of 
$\tau^- \rightarrow \pi^- \eta'(958) \nu_{\tau}$ should be less than $1.4 \times 10^{-6}$ \cite{etalimit}.

\section{$\tau^- \rightarrow h^- h^+ h^- \nu_{\tau}$ invariant mass spectra}

For the event selection, a sample of $\tau^- \rightarrow h^-h^+ h^- \nu_{\tau}$ decays 
events is selected by requiring the partner $\tau^+$ to decay leptonically. 
Within this sample, each of the mesons is uniquely identified as a charged 
pion or kaon, and the decay categorized as 
$\tau^- \rightarrow \pi^-\pi^+ \pi^- \nu_{\tau}$, $\tau^- \rightarrow K^-\pi^+ \pi^- \nu_{\tau}$, 
$\tau^- \rightarrow K^-K^+ \pi^- \nu_{\tau}$ or $\tau^- \rightarrow K^-K^+ K^- \nu_{\tau}$
where events with $K^0_S$ have been excluded. After events are selected the invariant mass distributions 
are analyzed. The $\tau^- \rightarrow h^-h^+ h^- \nu_{\tau}$ backgrounds between the channels caused 
by particle mis-identification, referred to as cross-feed throughout this paper, are normalized to the 
measured branching fractions in BaBar. 
The cross-feed backgrounds are estimated to be $(0.85\pm0.01)$\% 
for the $\pi^-\pi^+\pi^-$  channel, $(38.5\pm0.2)$\% for the $K^-\pi^+\pi^-$ channel, $(2.9\pm0.1)$\% for the 
$K^-K^+\pi^-$  channel and $(27.7\pm3.0)$\% for the $K^-K^+K^-$ 
channel, where the uncertainties are from MC statistics. 
The background fractions from events with an extra $\pi^0$ in the candidate samples are estimated 
to be $(3.6\pm0.3)$\% from $\tau^- \rightarrow \pi^-\pi^+ \pi^- \pi^0 \nu_{\tau}$ in $\tau^- \rightarrow \pi^-\pi^+ \pi^- \nu_{\tau}$, 
$(2.3\pm0.4)$\% from $\tau^- \rightarrow K^-\pi^+ \pi^- \pi^0 \nu_{\tau}$
in $\tau^- \rightarrow K^-\pi^+ \pi^- \nu_{\tau}$ , $(0.4\pm0.1)$\% from $\tau^- \rightarrow K^-K^+ \pi^- \pi^0 \nu_{\tau}$ in 
$\tau^- \rightarrow K^-K^+ \pi^- \nu_{\tau}$ and less than 5.0\% from 
$\tau^- \rightarrow K^- K^+ K^- \pi^0 \nu_{\tau}$ in $\tau^- \rightarrow K^- K^+ K^- \nu_{\tau}$. 
The non-$\tau$ backgrounds amount to less than 0.5\% of the
events for each channel. 
An arithmetic subtraction is used to remove the backgrounds 
from the invariant mass distributions for each channel. 
Detector effects are then removed using Bayesian Unfolding \cite{tauinvmunf}, which has been trained using the signal MC 
for each decay mode. An efficiency correction, initially obtained from MC and corrected using data control 
samples, is used to correct for efficiency losses from the event selection for each bin in the invariant mass distribution. 
When the statistical uncertainty on the MC is below 10\% the efficiency is determined using the neighbouring bins. 
The invariant mass distributions are then normalized to unity. The two-dimensional Dalitz distributions for 
slices of the three-body invariant mass are background subtracted and then efficiency-corrected. The bin width
for the Dalitz plots is chosen to be 25 MeV which is approximately 5 times the resolution of the two-body invariant masses.
The systematic uncertainties in this work that are taken into account include: the MC signal and 
background statistics, the potential biases resulting from the Bayesian Unfolding, the uncertainties related to 
particle identification, the uncertainties in the modeling of the EMC and tracking response, the modeling of 
the trigger, the luminosity and the modeling of the backgrounds. 

\begin{figure}[ht!] 
\begin{center}
\includegraphics[width=.48\textwidth]{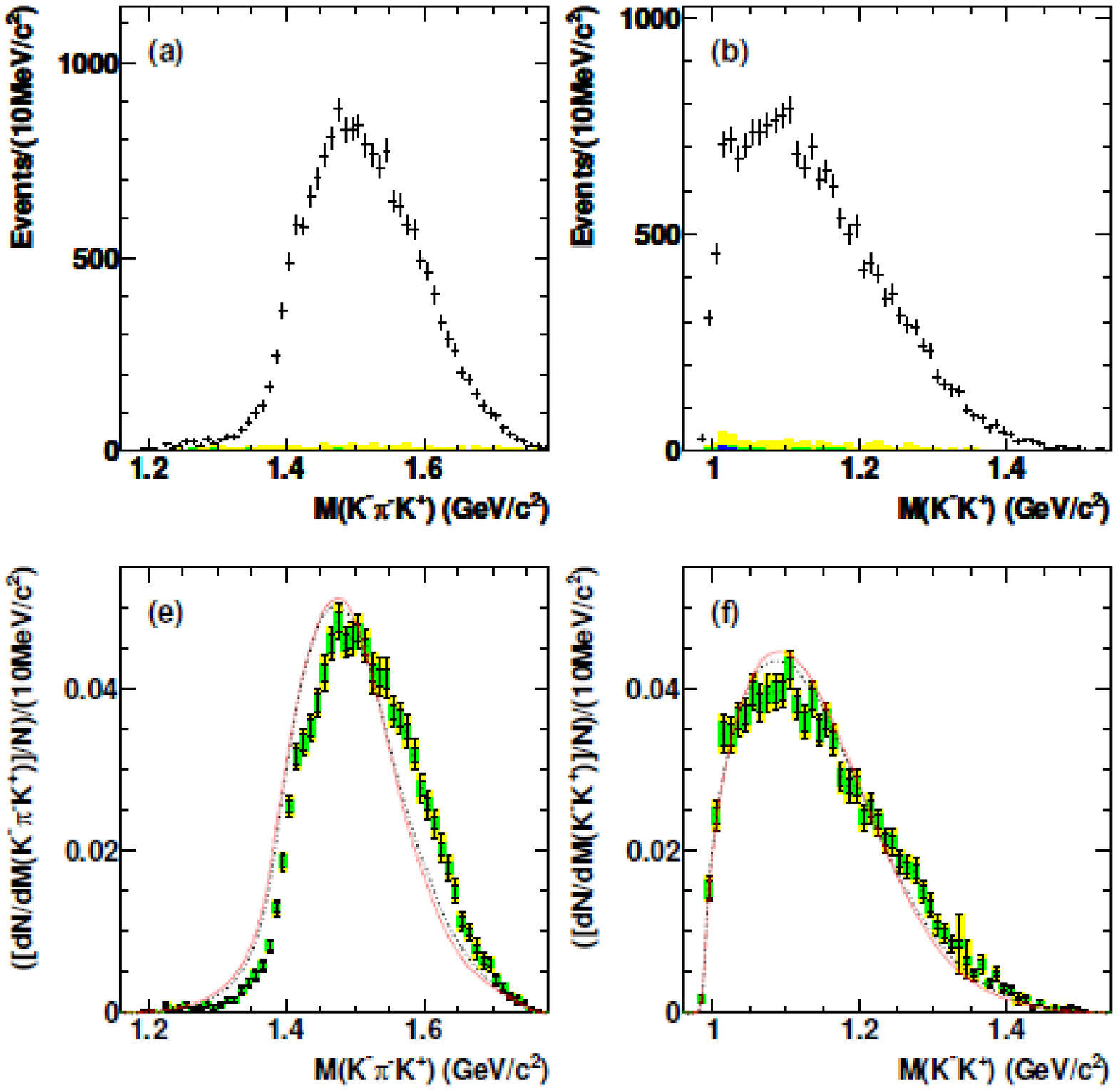}
\quad
\includegraphics[width=.48\textwidth]{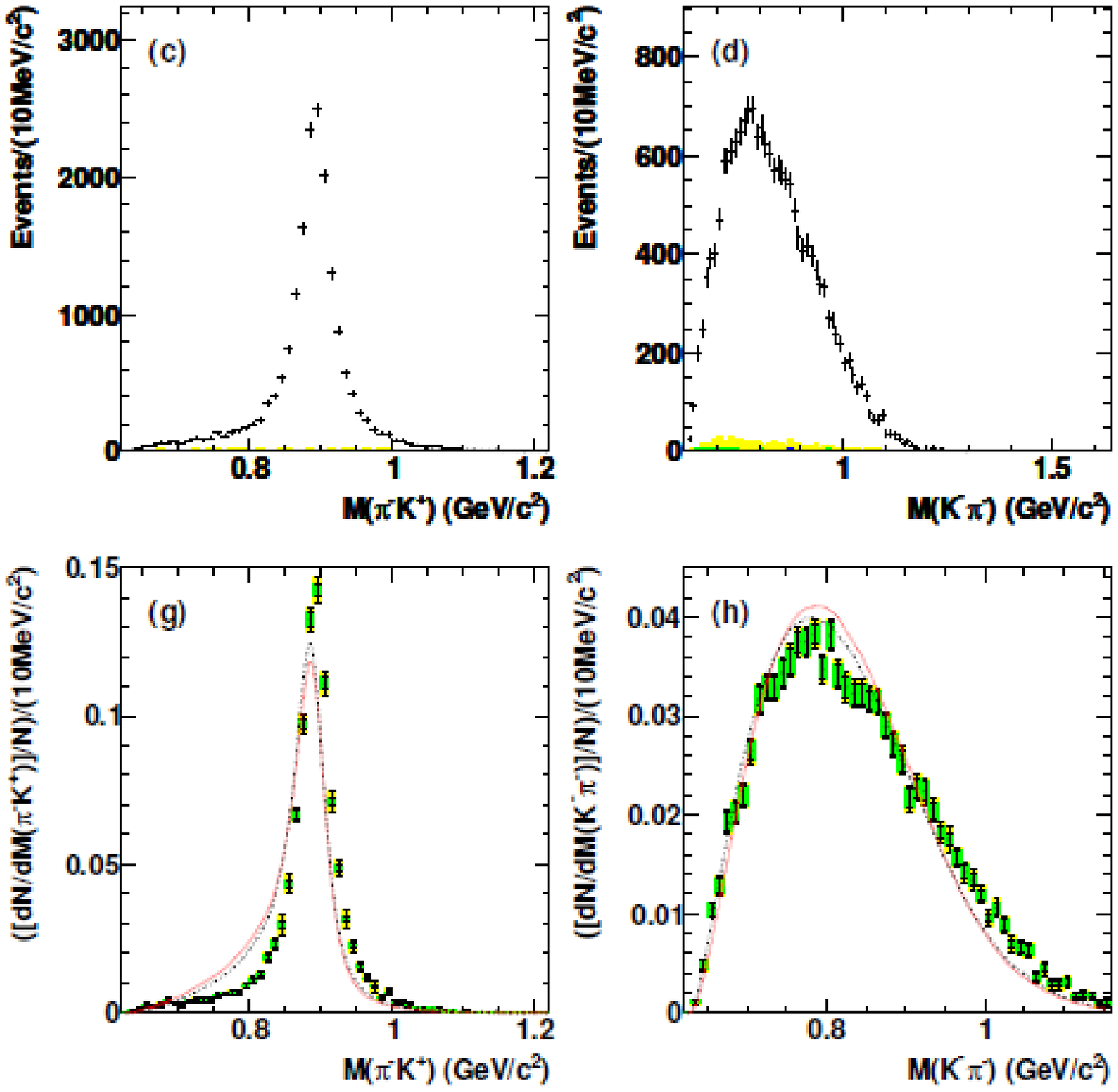}
\label{fig:pikk_spectra} 
\caption{\scriptsize The reconstructed and unfolded invariant mass spectra for the 
$\tau^- \rightarrow K^-\pi^+ \pi^- \nu_{\tau}$ channel. We present the reconstructed
invariant mass distributions for (a) $M(K^-\pi^-K^+)$, (b) $M(K^-K^+)$, (c) $M(\pi^-K^+)$ and (d) $M(K^-\pi^-)$ 
and the unfolded invariant mass spectra (e) $M(K^-\pi^-K^+)$, (f) $M(K^-K^+)$, (g) $M(\pi^-K^+)$ and (h) $M(K^-\pi^-)$. 
For the reconstructed mass plots, the data 
is represented by the points with the error bars representing the statistical uncertainty. 
The blue (dark) histogram represents 
the non-$\tau$ background MC, the green (medium dark) histogram represents the $\tau$ backgrounds excluding the 
$\tau^- \rightarrow K^-\pi^+ \pi^- \nu_{\tau}$ cross-feed which are represented by the yellow (light) histogram. 
For the unfolded mass plots, the data is represented by the 
points with the inner error bars (green) representing the statistical uncertainty and the outer error bars (yellow) representing
the statistical and systematic uncertainties. The integral of the unfolded distribution is normalized to unity. 
The black dashed line is the generator level MC distribution used in the BaBar simulation. The red dotted line is the 
CLEO tune for Tauola 2.8 \cite{tauola2}.}
\end{center}
\end{figure}


For the $\tau^- \rightarrow \pi^-\pi^+ \pi^- \nu_{\tau}$ it can be seen that
invariant mass $M(\pi^-\pi^+ \pi^-)$ is consistent with previous measurements \cite{pipipi}. 
The width of the $a_1(1260)$ observed in this study is larger than in Tauola \cite{tauola2}. 
It can also be seen that the $a_1(1260)$, which primarily decays through $\pi \rho$, is distorted 
by phase space constraints and in particular it can be seen that there is a large discrepancy 
between the data and the models in the low $M(\pi^+ \pi^-)$ region. The uncertainties of the
unfolded $\tau^- \rightarrow \pi^-\pi^+ \pi^- \nu_{\tau}$ spectra are limited by the 
particle-id and the relative background fraction. 

The $M(K^-\pi^-\pi^+)$ invariant mass distribution for
$\tau^- \rightarrow K^-\pi^+ \pi^- \nu_{\tau}$ has been measured previously by the OPAL, ALEPH, Belle \cite{pipipi}
and CLEO collaborations, 
where all but the latter unfolded the invariant mass spectrum. The LEP experiments and CLEO had limited 
statistics and are consistent both with the Belle results and the results presented here. 
However, the Belle results \cite{pipipi} for the invariant mass distribution and the results presented here are 
inconsistent. The discrepancy is most pronounced in the 1.4-1.7 GeV/c$^2$ range where the 
$\tau^- \rightarrow \pi^-\pi^+ \pi^- \nu_{\tau}$ background dominates, 
indicating that the difference is probably related to the estimate of this background. 
This hypothesis is consistent with the observed 
discrepancies in the $\tau^- \rightarrow \pi^-\pi^+ \pi^- \nu_{\tau}$ branching fraction. 
The channel $\tau^- \rightarrow \pi^-\pi^+ \pi^- \nu_{\tau}$ is observed to decay primarily through 
the $K_1(1270)$ and $K_1(1400)$ resonances, and then subsequently 
through the intermediate states $K*(892)\pi^-$ and $\rho^0 K^-$. 
This general decay structure is consistent with the measurements in \cite{kpipi}. 
It can also be observed that the $\rho^0$ which primarily comes from the $K_1(1270)$ is 
strongly constrained by phase space. 

The $M(K^-K^+ \pi^-)$ invariant mass distribution for $\tau^- \rightarrow K^- K^+ \pi^- \nu_{\tau}$ 
is consistent with the distribution in \cite{pipipi} for most of the invariant mass spectrum 
except for a slight excess in the region, 1.65-1.75 GeV/c$^2$.
This is where the Belle measurement has a large relative cross-feed contribution \cite{pipipi}. 
The primary decay mechanism is observed to be through $K*(892)K^-$. 
In contrast to the $\tau^- \rightarrow \pi^-\pi^+ \pi^- \nu_{\tau}$ 
and $\tau^- \rightarrow K^-\pi^+ \pi^- \nu_{\tau}$ channels, the $\tau^- \rightarrow K^- K^+ \pi^- \nu_{\tau}$
spectra are statistically limited and therefore do not have as strong of correlation as the 
aforementioned spectra have. However, there is still a dependence of the $K*(892)$ resonance on 
the phase space due to the normalization.

The $\tau^- \rightarrow K^- K^+ K^- \nu_{\tau}$ decay has been measured by 
Belle \cite{pipipi} and by BaBar \cite{kkkbbr}. The $M(K^-K^+)$ invariant mass was shown 
to decay predominantly through the $\phi$ resonance and 
with an upper limit of $B(\tau^- \rightarrow K^- K^+ K^- \nu_{\tau})< 2.5 \times 10^{-6}$ at 90\% CL \cite{kkkbbr}. 
The shape of the $M(K^-K^-K^+)$ distribution is consistent with the only other measurement \cite{pipipi}, 
however, the branching fractions measured by BaBar \cite{kkkbbr} and Belle \cite{pipipi} 
are inconsistent by more than 5.4$\sigma$.




\begin{thebibliography}{99}
\bibitem{ronban} S. Banerjee, B. Pietrzyk, J.M. Roney, Z. Was, {\it Phys. Rev. }{\bf D 77}, 054012 (2008). 
\bibitem{kos} J. P. Lees {\it et al.} (BABAR Collaboration), {\it Phys. Rev.} {\bf D 85}, 031102 (2012). 
\bibitem{kos2} J. Beringer {\it et al.} (Particle Data Group), {\it Phys. Rev.} {\bf D 86}, 010001 (2012).
\bibitem{hmtau} J.P. Lees {\it et al.} (BABAR Collaboration), {\it Phys. Rev. }{\bf D 86}, 092010, (2013).
\bibitem{tauinvmcab} N. Cabibbo, {\it Phys. Rev. Lett.}, {\bf 10}, 531–533, (1963).
\bibitem{tauinvmstrange} E. Gamiz, M. Jamin, A. Pich, J. Prades, F. Schwab, {\it Phys. Rev. Lett.}, {\bf 94}, 011803, (2005).
\bibitem{tauinvmunf} G. D'Agostini, {\it Nucl. Instrum. Meth.}, {\bf A362}, 487–498, (1995).
\bibitem{babar} B. Aubert {\it et al.} (BABAR Collaboration), {\it Nucl. Instr. Methods Phys. Res., Sect.} {\bf A 479}, 1 (2002).
\bibitem{babardet} B. Aubert {\it et al.} (BABAR Collaboration), {\it Phys. Rev. Lett.} {\bf 99}, 021603 (2007). 
\bibitem{kk2f} B. F.Ward, S. Jadach, Z.Was, {\it Comput. Phys. Commun.} {\bf 130}, 260 (2000).
\bibitem{tauola} S. Jadach, Z. Was, R. Decker J. H. Kuhn, {\it Comput. Phys. Commun.} {\bf 76}, 361 (1993).
\bibitem{jetset} T. Sjostrand, {\it Comput. Phys. Commun.} {\bf 82}, 74 (1994). 
\bibitem{photos} E. Barberio and Z. Was, {\it Comput. Phys. Commun.} {\bf 79}, 291 (1994). 
\bibitem{geant} S. Agostinelli {\it et al.} (GEANT4 Collaboration), {\it Nucl. Instr. Methods Phys. Res.}, Sect. {\bf A 599}, 250 (2003). 

\bibitem{cleoeta} A. Anastassov {\it et al.} (CLEO Collaboration), {\it Phys. Rev. Lett.} {\bf 86}, 4467 (2001). 
\bibitem{f1_decay} M. Acciarri {\it et al.} (L3 Collaboration), {\it Phys. Lett.} {\bf B 501}, 1 (2001).
\bibitem{f1_ratio} J. Beringer {\it et al.} (Particle Data Group), {\it Phys. Rev.} {\bf D 86}, 010001 (2012). 
\bibitem{etalimit} S. Nussinov, A. Soffer, {\it Phys. Rev.} {\bf D 80}, 033010 (2009). 

\bibitem{pipipi} M. Lee {\it et al.}, {\it Phys. Rev.}, {\bf D81}, 113007, (2010). 
\bibitem{tauola2} P. Golonka, B. Kersevan, T. Pierzchala, E. Richter-Was, Z. Was, {\it et al.}, {\it 
Comput. Phys. Commun.}, {\bf 174}, 818–835, (2006). 
\bibitem{kpipi} G. Abbiendi {\it et al.}, {\it Eur.Phys.J.}, {\bf C13}, 197–212, (2000). 
\bibitem{kkkbbr} B. Aubert {\it et al.}, {\it Phys. Rev. Lett.}, {\bf 100}, 011801, (2008).
 \end{thebibliography}
\end{document}